\documentclass[12pt]{article}
\usepackage[a4paper, margin=1in]{geometry}
\usepackage{graphicx}
\usepackage{hyperref}
\usepackage{amsmath}
\usepackage{amsfonts}
\usepackage{amssymb}
\usepackage{setspace}
\usepackage{lipsum}
\usepackage{xcolor}
\usepackage{listings}

\usepackage{authblk} % For author affiliations
\usepackage{geometry} % Adjust margins if necessary
\title{BACTA-GPT: An AI-Based Bayesian Adaptive Clinical Trial Architect}
\author[1]{Krishna Padmanabhan}
\author[1,2]{Danny Baker}
\affil[1]{Cytel, 675 Massachusetts Ave, Cambridge, MA 02139\\ \texttt{krishna.padmanabhan@cytel.com}}
\affil[2]{University of Connecticut, School of Computing, 371 Fairfield Way, Storrs, CT 06268\\ \texttt{daniel.baker@uconn.edu}}

\date{\today}

\begin{document}
\maketitle
\begin{abstract} 
\textbf{Background} Bayesian adaptive clinical trials offer a flexible and efficient alternative to traditional fixed-design trials, but their implementation is often hindered by the complexity of Bayesian computations and the need for advanced statistical programming expertise. The authors introduce a custom fine-tuned LLM designed to assist with this and lower barriers to adoption of Bayesian methods for adaptive clinical trials. \\
\textbf{Methods} This paper describes the development and fine-tuning of BACTA-GPT, a Large Language Model (LLM)-based tool designed to assist in the implementation of Bayesian Adaptive Clinical Trials. This engine uses GPT-3.5 as the underlying model and takes in Natural Language input from the Statistician or the Trialist. \\
\textbf{Results} The fine-tuned model demonstrates a viable proof-of-concept in its objectives. Test case evaluations show that the model is capable of generating a fit-for-purpose Bayesian model for an adaptive trial and evaluate its operating characteristics via simulations using \texttt{R} and \texttt{JAGS}.\\
\textbf{Conclusions} The integration of AI code generation has significant potential to lower technical barriers for the design and implementation of Bayesian Adaptive trials. But they also require attention to important considerations regarding validation and quality control.\\
\end{abstract}


\newpage 
\section{Introduction}

Clinical trials represent the cornerstone of evidence-based medicine, yet traditional fixed-design trials often face limitations in their ability to adapt to emerging data and evolving understanding of treatments. Adaptive clinical trials \cite{bhatt2016} have become an attractive alternative, offering flexibility and efficiency in trial design while maintaining scientific rigor. Many of these trials leverage Bayesian statistical methods \cite{lee2012} to update probabilities as new data becomes available, allowing for dynamic modifications to trial parameters and potentially reducing both time and resource requirements. \\ 

However, the implementation of Bayesian adaptive trials presents significant challenges, particularly in the realm of statistical programming and simulation. The complexity of Bayesian computations, coupled with the need for sophisticated simulation models, often requires extensive coding expertise and computational resources. This technical barrier can limit the adoption of adaptive designs, even when they may be the optimal choice for a given research question \cite{medicaloutreach2022}. \\

Large Language Models (LLMs) have recently emerged as powerful tools for code generation and automation across various domains \cite{chatgpt}. Their ability to understand context, generate accurate code, and adapt to specific programming requirements makes them particularly well-suited for statistical programming tasks. In the context of Bayesian adaptive trials, LLMs offer the potential to bridge the gap between complex statistical methodology and practical implementation. \\

The intersection of Bayesian statistics and artificial intelligence based coding agents presents a unique opportunity to enhance the accessibility and efficiency of adaptive clinical trial design. By leveraging LLMs' capabilities in natural language processing and code generation, we can potentially democratize access to sophisticated trial designs while maintaining the mathematical rigor required for clinical research. In this paper, we introduce BACTA-GPT, a custom fine-tuned LLM that is capable of assisting with code for Bayesian Adaptive Clinical trials in the R programming language, based on Natural Language input. We describe the process of fine-tuning and the prompts that were used to generate this model. We also demonstrate performance on held-out test cases.

\subsection{Literature Review}
\subsubsection{Bayesian Adaptive Trials}

The evolution of Bayesian adaptive trials in clinical research spans several decades, marked by theoretical developments, computational advances, and increasing practical implementation. Early work in the 1980s laid the foundation for modern adaptive trial designs, demonstrating the potential for interim analyses and dynamic sample size adjustment \cite{berry2006}. These theoretical frameworks have since been expanded and refined, leading to a rich ecosystem of methodological approaches. Bayesian methods excel in adaptive trials through their probabilistic framework that updates allocation probabilities toward better-performing treatments, enables natural interim stopping rules using posterior probabilities, and incorporates historical data through hierarchical borrowing. This creates a flexible yet rigorous foundation for modern adaptive trial design  \cite{lee2012}. \\

Significant real-world applications of Bayesian adaptive designs include the I-SPY2 trial \cite{barker2009} in breast cancer research, the STAMPEDE trial in prostate cancer \cite{james2008}, platform trials during the COVID-19 pandemic, such as REMAP-CAP \cite{angus2020}. The statistical implementation of Bayesian adaptive trials has been facilitated by various software tools:
\begin{itemize}
\item WinBUGS \cite{lunn2000}, one of the first software packages developed for Bayesian statistical analysis using Markov Chain Monte Carlo (MCMC) methods.

\item JAGS (Just Another Gibbs Sampler) \cite{plummer2003} has become a standard tool for Bayesian inference, offering flexibility in model specification and efficient MCMC sampling.

\item Stan \cite{carpenter2017} provides a more modern platform for Bayesian inference, with advantages in handling complex hierarchical models and improved convergence properties.

\item Commercial solutions like EastBayes \cite{cytel2021} and FACTS \cite{facts2023} have emerged to provide user-friendly interfaces for design and simulation.
\end{itemize}

Despite these advances, several challenges persist in the implementation of Bayesian adaptive trials:

\begin{itemize}

\item Technical Expertise Requirements: The implementation of Bayesian adaptive designs often requires specialized knowledge in both statistical theory and programming, creating a barrier to wider adoption.

\item Computational Complexity: The need for extensive simulations to validate operating characteristics can strain computational resources, particularly for complex adaptive designs.

\item Regulatory Considerations: While agencies like the FDA have become more accepting of Bayesian methods, they are still considered less well understood by stakeholders. 
\end{itemize}

The technical and computational challenges motivated the authors to consider using LLMs for their code-generating ability to simplify the process of implementing statistical models for Bayesian Adaptive Trials. It is also noteworthy that the US FDA is expected to release a draft guidance document in 2025 on the ``Use of Bayesian Methodology in Clinical Trials of Drug and Biological Products" \cite{fda2024}. This potentially points to a growing acceptance of Bayesian designs in randomized clinical trials by regulators.

\subsubsection{Fine-Tuning Large Language Models (LLMs)}

Fundamentally, the goal of an LLM is to model language. These models are useful for a wide range of tasks including summarization, Q\&A, translation, text generation, text sentiment analysis, and, since code follows a similar structure to spoken language, code generation. In order to have a solid grasp on the patterns in natural language – both semantic and contextual – frontier LLM models are trained on immense amounts of data and contain hundreds of billions of parameters \cite{deepseek, chatgpt, llama}. Conditioned on a question as textual context, LLMs seek to predict the next most likely word in an autoregressive manner. That is, they generate responses by predicting each word sequentially while considering the given question and the preceding text. We can compare this prediction to a ground-truth correct answer and optimize the model’s parameters to improve accuracy. In this way, LLMs learn the distribution of human language. \\

LLMs often undergo another stage of training called fine-tuning in which they are specifically designed to handle Q\&A style chats. For example, ChatGPT is an LLM specifically fine-tuned to function as a chat bot \cite{chatgpt}. This is fueled by by the Transformer architecture \cite{vaswani2023} which allows for the efficient processing of queries as well as a larger context window. Originally trained on general texts, ChatGPT is then fine-tuned to use its learned understanding of human language in the form of Q\&A. This fine-tuning stage allows developers to incorporate specific patterns and objectives into their models.\\

Prompt engineering is a strategy used to further improve the performance of trained LLMs by creating better input prompts that produce higher quality results. Two powerful techniques are few-shot learning and Chain-of-Thought reasoning. In few-shot learning, users effectively model desired input-output patterns by providing the LLM with a few examples of input-output pairs at inference time. This way, when providing the actual query, the model has strong examples in context history. Typically, few-shot learning consists of 10-100 strong examples while one-shot learning consists of one strong example and zero-shot learning consists of prompting the LLM without any guidance. This strategy does not require training and can greatly reduce the need for a task-specific dataset. \cite{fewshot}. However, it is not as powerful as fine-tuning because the underlying model is not changed. \\

Chain-of Thought reasoning (CoT) is a technique influenced by the human nature to reason through a problem and arrive at the correct answer. For example, solving a complex word problem requires step-by-step reasoning rather than immediate recall of the correct answer. This step-by-step approach is referred to as a CoT and can be leveraged to improve LLM performance. One study found that modeling CoT in a few-shot learning approach more than doubled the accuracy of the GSM8K benchmark \cite{CoT}. CoT is beneficial for a few reasons. First, it allows the model to decompose problems into smaller steps and assert more computational power to the more difficult steps. Second, the reasoning steps offer the ability to interpret the model’s thought process. Third, this technique does not require training, and can therefor be used with off-the-shelf models. Recently there has been an increased focus on training models using CoT examples to further improve reasoning skills \cite{deepseek, chatgpto1}. While prompt engineering is a very useful approach to eliciting better responses, it is not a perfect solution. In such cases it is necessary to fine-tune a model to a specific goal.\\

Since LLMs are designed to understand language, they can effectively learn programming languages and complete various coding tasks. This advantage allows for great improvements in efficiency for both software developers and any professional who uses programming in their daily work. To motivate the development of this skill, coding benchmarks exist to mark the progress of LLMs and their capabilities. For example, SWE-bench Verified \cite{swebench} was developed to determine how well LLMs can resolve real-world issues from GitHub. Similarly, CodeElo \cite{codeelo} introduces a programming benchmark (derived from CodeForces\footnote{https://codeforces.com}) that closely mimics both the question and human-evaluation styles of coding competitions. Although not strictly code-related, the AIME 2024 benchmark \cite{aime2024} seeks to evaluate an LLMs ability to perform mathematical reasoning which is highly adjacent to the goals of our paper. Currently DeepSeek-R1 \cite{deepseek} is performs the best on AIME 2024 and SWE-bench Verified with a performance of 79.8\% and 49.2\% respectively. Meanwhile, OpenAI-o1 \cite{chatgpto1} performs the best on CodeElo with an accuracy of 96.3\% Significant effort is being done to push the boundary of coding abilities and outperform these existing state of the art methods.

\subsection{Advantages of Using LLMs for Bayesian Trial Development}

The use of LLMs in helping generate Bayesian adaptive trial models may offer distinct advantages that address longstanding challenges in the field. At the technical level, LLMs dramatically reduce coding effort through automated generation of R code for trial designs, seamless translation of statistical specifications into executable code, and built-in validation of code structure and syntax. This automation extends to generating comprehensive documentation and integrating with existing statistical software packages, while also providing a natural language interface for trial design specification and interactive debugging support.\\

The practical benefits of LLMs manifest in their ability to democratize access to complex trial designs through improved accessibility and iterative optimization capabilities. By offering code templates for common design patterns and simplified modification of existing trials, LLMs reduce the need for specialized programming expertise. They enable rapid prototyping of different design options, automate sensitivity analyses, and generate comprehensive simulation scenarios while optimizing code for computational efficiency and incorporating best practices in Bayesian computation.\\

These advantages are amplified by the enhanced speed and scalability that LLMs bring to trial implementation. Through code generation, parallelization and efficient handling of complex adaptive rules, LLMs streamline the process of conducting large-scale simulation studies. Their ability to integrate with high-performance computing environments and handle large datasets positions them as powerful tools for modern clinical trial design. 

\section{Materials and Methods}
\subsection{Development of BACTA-GPT: Workflow}
The goal of BACTA-GPT is to leverage the coding ability of LLMs to design Bayesian adaptive trails. Specifically, we build BACTA-GPT on ChatGPT 3.5 \cite{chatgpt35} and aim to take advantage of its strong code-generation capabilities. Given a detailed overview of the desired trial from a statistician or the trialist, he objective was for BACTA-GPT to be able to comprehend the problem, design a Bayesian hierarchical model, and produce code to run a simulation of the trial. To achieve this goal, we carefully designed a workflow inspired by CoT reasoning that optimizes results. Rather than asking ChatGPT to generate the entire trial all at once, a step-by-step, incremental approach enhances both comprehension and performance. The workflow designed is as follows. 
\begin{enumerate}

\item \textbf{Trial Parameters:} It is first necessary to provide the LLM with the trial information including objectives, endpoints, priors, and even a model structure suggestion. This allows the model a chance to distill the user’s information. Injecting any domain-expert knowledge here is crucial because it guides the LLM in a productive direction. BACTA-GPT will return a summary of the problem, goals, and proposed model design.
\item \textbf{Model Formulation:} This stage describes the trial design mathematically by defining the priors and hyperparameters. 
\item \textbf{JAGS Model:} The next stage is the generation of the model - the authors chose the JAGS approach, driven by their familiarity and comfort with the framework. This can be seen as a translation problem where we go from a natural language/mathematical description of the trial to more rigid code representation of the trial.
\item \textbf{Full Trial - single iteration: } In this step, we ask the model to place the model in the context of a full iteration of a trial. A key step here is for the model to generate a dataset that mimics the trial and is compatible with the generated Bayesian model. This step ensures the model's understanding of the trial data structure, intersection with the Bayesian hierarchical model and the objectives of the interim and final analyses.
\item \textbf{Simulation:} In this stage, BACTA-GPT develops the full simulation by generating the code needed to run a full adaptive trial. This includes multiple runs of generating individual trial results based on parameters provided, running analyses based on the JAGS model, and adapting the trial arms based on any intermediate posterior checks. This stage also provides code to produce posterior predictive checks as well as additional output summaries of the trial. 
\end{enumerate}

We found that breaking down the task into specific deliverables dramatically improved BACTA-GPT's ability to produce meaningful results. This improvement is likely attributed to the use of more specific, goal-oriented prompts and the large context window in ChatGPT 3.5. \\ 

We trained BACTA-GPT on many different trial types to create a more robust understanding of clinical trial design and improve generalization. These types include models using Linear Regression, Logistic Regression, Meta Analyses, GLMs (Gamma, Beta and Poisson regression), Regression with categorical variables, Regression with Over-dispersed outcomes, Random effects models, Mixture models, and Survival Analysis.  All of our examples -- both training and test -- demonstrate the same workflow.\\

Each stage offers users a chance to analyze the model’s understanding and make any corrections as necessary. For example, if BACTA-GPT misses some critical information in the model definition, statisticians or trialists can critique the model to ensure the output is correct for each subsequent step. Given the complexity of this task, this approach is considerably more reliable than simply asking for the full simulation at once where finite computational power needs to be allocated to more tasks and errors at the beginning of the response are carried through. This technique of breaking down the trial into a step-by-step development ensures that each intermittent stage is correct, and the full trial is created from accurate components. \\

\subsection{Fine-Tuning Process and Optimization}

Original attempts at developing trials with OpenAI’s ChatGPT 3.5-Turbo involved a few-shot CoT prompting approach similar to that of \cite{CoT}. This approach involved incorporating multiple examples of the desired workflow into the model's chat history before posing the actual query. The full examples demonstrated the development of a trial step by step. It was essential to include well-structured responses with thorough definitions for priors, hyperparameters, and clean code. However, despite multiple rounds of prompt tuning and refinement, few-shot CoT prompting presented challenges. ChatGPT had difficulty generating precise prior definitions and JAGS code that strictly adhered to the model specification.\\

These shortcomings were exacerbated in more complex examples or when dealing with compound conditions. This is primarily due to the inherent complexity of these problems and high reasoning capabilities needed to solve them. Another limitation is the long nature of our workflow combined with the constraints of ChatGPT’s context window. ChatGPT 3.5-Turbo’s 16,385 token context window truncates the entire chat history to only the previous 16,385 tokens (around 12,000 words), including both the user prompts and model responses. While this context window is sufficiently large to include a few examples, we were unable to achieve reliable, consistent results. \\ 

To further improve and refine model performance, fine-tuning was used to update the model’s parameters directly. This powerful technique allowed us to create a version of ChatGPT 3.5-Turbo that specializes in the design of Bayesian adaptive clinical trials which we refer to as BACTA-GPT. To do this, we manually crafted a small set of 32 separate Bayesian adaptive clinical trials. This training set consisted of different hierarchical models. These were also completely reworded and organized as a data augmentation technique to create multiple further copies, with an aim to enhance the model’s robustness to variations in user input styles. All 32 examples were implemented as multi-turn chats that simulate the development of a trial from start to finish. \\

Leveraging OpenAI’s API for training, BACTA-GPT was trained using a cross-entropy loss function. To achieve the optimal performance for our limited training set we tuned hyperparameters such as batch size and training epochs. We found ideal performance training with a batch size of 1 and 3 epochs. Due to the length of each example, we found this batch size to be appropriate. Further hyperparameter tuning included adjusting the model’s temperature and Top-P sampling. Since lower temperatures are preferable for standardized tasks such as coding, optimal performance was achieved with a temperature of 0.3 and Top-P set to 1.

\section{Results}

To test BACTA-GPT, we created a test set of 4 unseen examples and developed full trials using our engineered workflow. Due to the computational demand of creating a training set for this task, the test examples did not have explicit ground truths to compare to. Instead, we employed a variety of manual and qualitative checks by subject-matter experts. For each test example, we manually scanned through the results ensuring that the responses were accurate, and the model was properly defined. We also ran the generated code and printed out posterior summaries to ensure reasonable properties. The accuracy of posterior quantities, posterior predictive checks and inferences were the main performance metrics we used to validate BACTA-GPT. Furthermore, we tested the full trial simulation including adaptive abilities and found that the full code generated was accurate and ran successfully.  Future work will incorporate more quantitative and robust metrics to evaluate BACTA-GPT’s ability to both interpret trial designs and generate accurate code. This requires larger, well-labeled datasets. A (held-out) test example for a trial measuring effect of a treatment on growth rates using non-linear regression model is given below. We include key elements of the final output here, but for brevity the full examples including intermediate outputs can be seen in the Appendix~\ref{appendix}. The series of prompts shown in the Appendix generally apply in the same way for any example for best results with BACTA-GPT. \newline\newline

% \textcolor{red}{KP: Unclear whether this is the training step. Or is this how the model interacts with the user when fully trained? Can you provide more context and be clearer about what step this represents and its objectives? IMHO, it is best to show the results of a test evaluation in order to demonstrate the capabilities for the fine tuned model. And the training prompt can be pushed out to an appendix. If you agree, then we can move this sub-section to the end of Section 6. And in this place, you can simply link to the appendix where we show a training prompt \\ }

\textbf{Key Trialist Prompts \\}

\noindent This section displays some of the key input prompts from the statistician or the trialist to the BACT-GPT system, providing the necessary details for the desired trial. Note that some prompts have been edited in minor fashion for clarity in this section, but have been presented verbatim in the Appendix. 

\begin{verbatim}
Trial Specifications:
    I have a continuous outcome variable Y, a growth rate, 
    one observation each from n child subjects in a clinical
    trial and this outcome is assumed to be normally 
    distributed with unknown variance.

Goal:
    Model Y using a regression with a two parameters: 
    a treatment variable X (0 or 1) and a 
    nonlinear function of the age (A), specifcally, a 
    parameter alpha raised to the power of age.

Requirements:
    Set uniform priors for alpha between 0 and 1.5
    For the sampling variance, please specify uniform prior 
    on log variance or the log sd scale. Set completely
    non informative prior for the coefficient of treatment.
    
Please provide the full JAGS code in R for this model?

Can you provide me with R code to generate a hypothetical 
simulated dataset for this trial, which is compatible with 
this JAGS model, by utilizing the assumptions below?

Set up the environment in R and load necessary 
libraries

Provide code to run this model and print out a 
summary of the posterior samples, based on the analysis of 
this dataset using the JAGS model.

Given this code, please help me change it such that the trial 
can be analyzed midway. 
Provide the R code to conduct an analysis on
this dataset with the JAGS model and calculate two separate
probabilities. First, the probability that the treatment
effect is > 10 units. Second, the probability that the
treatment effect is > 5 units. 

These probabilities will be used in order to make decisions 
at the IA regarding the future course of the trial (Early Stop).
If the trial is stopped at the IA (for efficacy or futility),
conduct final analysis. Else, add the necessary number of patients 
and preceed to final analysis as planned. 

At the final analysis for the trial, Provide R code to 
compute the posterior probability that the treatment 
difference is greater than 5 units. Declare trial Success or Failure 
based on this probability achieving a pre-defined threshold.


\end{verbatim}

\textbf{Key Sections of the BACTA-GPT Output \\}

\noindent This section displays some of the key output from the BACTA-GPT fine-tuned model in response to the trialist's input prompts above. Again, some comments accompanying the outputs (but not the outputs themselves)  have been edited in minor fashion for clarity here, but have been presented verbatim in the Appendix.

\begin{verbatim}
# JAGS Bayesian model for Primary Efficacy endpoint 

model_string <- "
model {
# Likelihood
for (i in 1:n) {
Y[i] ~ dnorm(mu[i], tau)
mu[i] = beta0 + beta1*X[i] + alpha^A[i]
}

# Priors for the model 
beta0 ~ dnorm(0, 1.0E-6)
beta1 ~ dnorm(0, 1.0E-6)
alpha ~ dunif(0, 1.5)
tau ~ dgamma(0.001, 0.001) # precision
sigma2 <- 1 / tau
}
"

# BACTA-GPT defined likelihood function

The response variable Y is normally distributed, so we can 
write the likelihood as:
Y[i] ~ Normal(mu[i], sigma2), where i = 1, 2, ..., n

The mean of the normal distribution, mu[i], is modeled as a
function of the treatment variable X and the nonlinear 
function of age:
mu[i] = beta0 + beta1 * X[i] + beta2 * alpha^A[i]

Here, beta0 is the intercept, beta1 is the coefficient for
the treatment variable, and beta2 is the coefficient for the
nonlinear function of age.

# Generation of Stage 1 dataset (N=200): 
Columns = Patient ID, Age, Treatment, Outcome
......  ..... (omitted data generation code for brevity here)

# The trial dataset
  trial_data <- data.frame(
    Y = Y,
    X = X,
    A = age)
  
# Prepare the data for JAGS
jags_data <- list(
  Y = trial_data$Y,
  X = trial_data$X,
  A = trial_data$A,
  n = nrow(trial_data)
)

# Initialize the model
jags_model <- jags.model(textConnection(model_string), 
                         data = jags_data, n.chains = 3)

# Burn-in
update(jags_model, 5000)

# Draw samples
mcmc_samples <- coda.samples(jags_model, 
                             variable.names = c("beta0", 
                             "beta1", "alpha", "sigma2"),
                             n.iter = 10000)
                             
# Extract the posterior samples
mcmc_list <- as.mcmc.list(mcmc_samples)

# Calculate the treatment difference for each iteration
treatment_diff <- as.matrix(mcmc_list[,'beta1'])

# Calculate the posterior probability that the treatment 
# difference is greater than delta
prob_treatment_diff_gt_10 <- mean(treatment_diff > 10)
prob_treatment_diff_gt_5 <- mean(treatment_diff > 5)

# Implement Interim Analysis (IA) Decision Rules
if (prob_treatment_diff_gt_10 > 0.95) {
  cat("Early Success\n")
} else if (prob_treatment_diff_gt_5 < 0.05) {
  cat("Early Futility\n")
} else {
# Generate the full dataset
  # Generate age, treatment and outcome for Stage 2 patients
......  ..... (omitted data generation code for brevity here)
# Create the full trial dataset
  trial_data <- data.frame(
    Y = Y,
    X = X,
    A = age)
  
# Calculate the posterior probability that the treatment 
# difference is greater than 5
prob_treatment_diff_gt_5 <- mean(treatment_diff > 5)
final_success <- prob_treatment_diff_gt_5 > 0.95
}

\end{verbatim}

\section{Discussion}

The above example demonstrates an example of our workflow that can be leveraged to develop a trial for a Non-linear Regression model to measure growth rates. The first step is to provide the model with the trial information, goals, and requirements. The model should distill the information into a concise summary providing statisticians and trialists with a chance to confirm that the trial is being interpreted correctly. The second step is to ask the model to define the parameter priors for this trial along with a brief motivation. The third step is to ask for the JAGS code. The model returns accurate JAGS code along with a description of the variables. In the fourth step, the user can provide information about a hypothetical dataset to run the trial on. Finally, step 5 allows the user to add adaptive elements based on intermediate results.\\

Additional queries could be made after these ones such as, ``\texttt{Can you please provide a model summary and give guidance on hyperparamter selection?}" and ``\texttt{Provide R code to calculate posterior probability that the treatment difference is > 0.}", as well as ``\texttt{Please provide R code to simulate the trial 1000 times."} These examples were excluded in the interest of brevity but were included in our training and validation process. We note that BACTA-GPT retains the impressive capabilities of standard LLMs, allowing it to answer diverse questions and incorporate feedback.\\

A key outcome of this work is the reduction of manual labor and time needed to develop a trial. Statisticians or trialists can use BACTA-GPT to handle some of the smaller tasks as well as prompt it for insights on more difficult questions. A short conversation with our model can significantly reduce the time required for model development, coding, debugging, and setting up an adaptive trial. When used as a tool, BACTA-GPT also offers the ability to check and validate work making the development of trials less error-prone.\\

Despite the successful proof-of-concept with BACTA-GPT, we identify a few persistent challenges and limitations. In some cases, human intervention is required to correct LLM understanding. This may look like prompting the model that it missed some crucial information, tweaking the input prompt wording, simply retrying the response generation, or manually modifying the generated code. When generalizing to unseen examples BACTA-GPT struggled to correctly define the model when the ask was complex.  This highlights the need for a more extensive training set to further improve generalization and robustness. This task is expensive and difficult due to the complexity of each problem but is critical to the advancement of clinical trial AI-assistants. We identify a source of bias stemming from our designed workflow. BACTA-GPT works best when adhering to this set of prompts but suffers when approaching problems from other angles or different prompting styles. While the designed workflow effectively maximized the utility of a our training set, the inclusion of further data augmentation and diverse prompting styles are needed for future improvements. \\

The authors plan to expand the scope of BACTA-GPT to support other common types of Bayesian Adaptive Clinical trials encountered in their consulting work. The next wave of priorities include:
    \begin{itemize}
        \item \textbf{Survival models} (using \textbf{INLA \cite{rue2016} or STAN}) for Bayesian survival analysis. This is driven by the fact the the baseline hazards have to be fully specified in a JAGS model, complicating the task beyond necessary, for most survival analysis tasks. 
        \item \textbf{Historical data borrowing} for improved decision-making. This includes approaches such as power prior \cite{chen2025} and Meta Analytic priors \cite{schmidli2014}. 

    \end{itemize}\
    
The authors also plan to engage a team of statisticians and trialists for "red-team" testing and improving BACTA-GPT on an invited basis.

\section{Conclusion}

The integration of AI code generation with Bayesian adaptive trial design has significant potential to benefit clinical research methodology. While AI-assisted approaches, particularly through LLMs, offer remarkable advantages in code generation and trial simulation, they also introduce important considerations for validation and implementation. The primary advantage lies in the rapid prototyping and implementation of complex adaptive designs that might otherwise require weeks of specialized programming effort. This acceleration of the design process, combined with AI's ability potentially go beyond the current use case and evaluate the outputs of its own code, can result in a paradigm shift. However, these benefits must be carefully weighed against the need for rigorous validation of AI-generated code and the maintenance of statistical integrity in trial design. \\

The democratization of complex trial simulations through AI tools represents a significant potential benefit for this use case. Historically, sophisticated Bayesian adaptive designs have been largely confined to well-resourced  institutions with specialized statistical programming expertise. AI-assisted tools are beginning to break down this barrier by providing intuitive interfaces and automated code generation that make advanced trial designs accessible to a broader range of researchers and institutions. This democratization could particularly benefit severe and rare disease research and in resource-limited settings, where adaptive designs could maximize the efficiency of limited patient populations and research budgets. However, this accessibility must be balanced with ensuring users maintain sufficient statistical understanding to appropriately interpret and validate the designs they implement. The future likely lies in hybrid approaches that combine AI's efficiency with human expertise in trial design and statistical theory, creating a more inclusive yet rigorous landscape for clinical research. \\

\noindent \textbf{Funding}
The work involved in the development of BACTA-GPT was conducted by the authors during a period when they were both paid employees of Cytel Inc. \\ 

\noindent \textbf{Conflict of Interest}
The Authors do not note any Conflicts of Interest. \\

\noindent \textbf{Author Contributions}
Both authors contributed meaningfully to the development of the manuscript and the underlying work.\\

\noindent \textbf{Supplemental Information} An appendix containing the full conversation by the Statistician with BACTA-GPT, in a Q$\&$A format, will be made available online. 

\newpage
\bibliographystyle{unsrt}
\bibliography{references} % Include a BibTeX file named "references.bib"
% \printbibliography
% \appendix
% \documentclass[12pt]{article}
% \usepackage[a4paper, margin=1in]{geometry}
% \usepackage{graphicx}
% \usepackage{hyperref}
% \usepackage{amsmath}
% \usepackage{amsfonts}
% \usepackage{amssymb}
% \usepackage{setspace}
% \usepackage{lipsum}
% \usepackage{xcolor}
% \usepackage{listings}
% \lstset{basicstyle=\ttfamily, 
%         breaklines=true,
%         breakatwhitespace=true,
%         columns=fullflexible,
%         frame=none,
%         xleftmargin=0pt}

% \title{BACTA-GPT: Appendix}
% \date{\today}
% \author{Krishna Padmanabhan, Danny Baker}

% \begin{document}

% \maketitle

\appendix
\section{Full Example}
\label{appendix}
\textbf{Statistician Prompt 1}
% \texttt{
\begin{verbatim}
I am a statistician designing a clinical trial using Bayesian 
Hierarchical modeling. Given the following trial, can you 
please do the following two steps.

Step 1: List the necessary data and model requirements.
Step 2: According to the necessary data and model 
requirements, mathematically define the likelihood function 
as a hierarchical model.

Trial Specifications:
    I have a continuous outcome variable Y, a growth rate, 
    one observation each from n child subjects in a clinical
    trial and this outcome is assumed to be normally 
    distributed with unknown variance.

Goal:
    Model Y using a simple linear regression with a two 
    parameters - a treatment variable X (0 or 1) and a 
    nonlinear function of the age (A), specifcally, a 
    parameter alpha raised to the power of age.

Requirements:
    Set uniform priors for alpha between 0 and 1.5
    For the sampling variance, please specify uniform prior 
    on log variance or log sd scale.
    Set completely non informative prior for the coefficient 
    of treatment.

    
\end{verbatim}
% }

\textbf{BACTA-GPT Response 1}
\begin{verbatim}
Step 1:
Data and requirements:
- Continuous outcome variable Y with one observation per 
  child subject in the clinical trial.
- Y is assumed to be normally distributed with unknown 
  variance.
- Simple linear regression model with two parameters: 
  treatment variable X (0 or 1) and a nonlinear function of 
  age (A) with parameter alpha raised to the power of age.
- Uniform prior for alpha between 0 and 1.5.
- Uniform prior for the sampling variance on the log variance 
  or log standard deviation scale.
- Completely non-informative prior for the treatment 
  coefficient.

Step 2: Define the likelihood function

The response variable Y is normally distributed, so we can 
write the likelihood as:
Y[i] ~ Normal(mu[i], sigma2), where i = 1, 2, ..., n

The mean of the normal distribution, mu[i], is modeled as a
function of the treatment variable X and the nonlinear 
function of age:
mu[i] = beta0 + beta1 * X[i] + beta2 * alpha^A[i]

Here, beta0 is the intercept, beta1 is the coefficient for
the treatment variable, and beta2 is the coefficient for the
nonlinear function of age.


\end{verbatim}

\textbf{Statistician Prompt 2}
\begin{verbatim}
Can you please mathematically define the priors for the 
parameters? and what is the motivation behind the 
hyperparameter selection?


\end{verbatim}

\textbf{BACTA-GPT Response 2}
\begin{verbatim}
For the intercept (beta0):
beta0 ~ Normal(0, 1000)

For the treatment coefficient (beta1):
beta1 ~ Normal(0, 1000)

For the coefficient of the nonlinear function of age (beta2):
beta2 ~ Normal(0, 1000)

For the nonlinear function parameter (alpha):
alpha ~ Uniform(0, 1.5)

For the sampling variance (sigma2):
sigma2 ~ Uniform(0, 1000)

The choice of a normal prior with a large variance for the
intercept, treatment coefficient, and nonlinear function 
coefficient is motivated by the desire to allow the data to
inform these parameters without imposing strong prior 
beliefs. The uniform prior for alpha between 0 and 1.5 is 
chosen to reflect the expected range of values for this 
parameter. The uniform prior for the sampling variance on the
log scale is a common choice for non-informative priors on 
variances.


\end{verbatim}

\textbf{Statistician Prompt 3}
\begin{verbatim}
Looks great. Can you please provide the full JAGS code 
in R for this model?


\end{verbatim}

\textbf{BACTA-GPT Response 3}
\begin{verbatim}

model_string <- "model {
# Likelihood
for (i in 1:n) {
Y[i] ~ dnorm(mu[i], tau)
mu[i] <- beta0 + beta1 * X[i] + beta2 * pow(alpha, A[i])
}

# Priors
beta0 ~ dnorm(0, 1.0E-3)
beta1 ~ dnorm(0, 1.0E-3)
beta2 ~ dnorm(0, 1.0E-3)
alpha ~ dunif(0, 1.5)
tau <- 1 / sigma2
sigma2 ~ dunif(0, 1.0E+3)
}"

The JAGS model code implements the likelihood and priors as 
defined earlier. The response variable Y is modeled as 
normally distributed, with the mean mu[i] depending on the 
treatment variable X and the nonlinear function of age. The 
priors for the intercept, treatment coefficient, and 
nonlinear function coefficient are set to normal 
distributions with mean 0 and precision 1.0E-3. The prior for
alpha is a uniform distribution between 0 and 1.5. The 
sampling variance (sigma2) is modeled on the log scale with a 
uniform prior.


\end{verbatim}

\textbf{Statistician Prompt 4}
\begin{verbatim}
Can you provide me with R code to generate a hypothetical 
simulated dataset for this trial, which is compatible with 
this JAGS model, by following these 4 steps which are in a 
bullet list with hashtag signs as bullets.

# Step1: Create code to generate a hypothetical clinical trial
dataset for this with 400 patients (1:1 randomization to 
active and placebo). The standard deviation of Y is 20 in 
both treatment groups and treatment effect should be about 6
units of difference (with higher mean for X =1 group). The 
age is randomly normally distributed across all patients with
a mean of 8 and a sd of 4 and truncate all observations of 
age < 3 to be equal to 3. 

# Step 2: Restate the JAGS model

# Step 3: Set up the environment in R and load necessary 
libraries

# Step 4: Provide code to run this model and print out a 
summary of the posterior samples, based on the analysis of 
this dataset using the JAGS model.


\end{verbatim}

\textbf{BACTA-GPT Response 4}
\begin{verbatim}
#Step 1: Generate the dataset
# Generate age
age <- rnorm(400, mean = 8, sd = 4)
age <- pmax(age, 3)  # Truncate age at 3

# Generate treatment and outcome
X <- rep(c(0, 1), each = 200)
alpha <- 1.1  # True value of alpha
A <- age - mean(age)  # Center age
mu <- 10 + 6*X + alpha^A
Y <- rnorm(400, mean = mu, sd = 20)

# Create the dataset
trial_data <- data.frame(
Y = Y,
X = X,
A = age
)

# Save the dataset
write.csv(trial_data, "clinical_trial_data.csv", 
          row.names = FALSE)

# Display summary statistics
summary(trial_data)

#Step 2: Restate the JAGS model
model_string <- "
model {
# Likelihood
for (i in 1:n) {
Y[i] ~ dnorm(mu[i], tau)
mu[i] = beta0 + beta1*X[i] + alpha^A[i]
}

# Priors
beta0 ~ dnorm(0, 1.0E-6)
beta1 ~ dnorm(0, 1.0E-6)
alpha ~ dunif(0, 1.5)
tau ~ dgamma(0.001, 0.001) # precision
sigma2 <- 1 / tau
}
"

#Step 3: Set up the environment and load libraries
library(rjags)
library(coda)

#Step 4: Run the model
# Load the data
trial_data <- read.csv("clinical_trial_data.csv")

# Prepare the data for JAGS
jags_data <- list(
Y = trial_data$Y,
X = trial_data$X,
A = trial_data$A,
n = nrow(trial_data)
)

# Initialize the model
jags_model <- jags.model(textConnection(model_string), 
                         data = jags_data, n.chains = 3)

# Burn-in
update(jags_model, 5000)

# Draw samples
mcmc_samples <- coda.samples(jags_model, 
                             variable.names = 
                             c("beta0", "beta1", "alpha", 
                             "sigma2"), n.iter = 10000)

# Summarize the results
print(summary(mcmc_samples))


\end{verbatim}

\textbf{Statistician Prompt 5}
\begin{verbatim}
ADAPTIVE ELEMENTS:
Given that code, please help me change it such that the trial 
can be analyzed midway. First, rather than running a trial 
for all 400 patients, please first generate data for 100 
patients in each group (total = 200, prior to the interim 
analysis) according to the same data specifications as 
before. Then provide the R code to conduct an analysis on 
this dataset with the JAGS model and calculate two separate 
probabilities. First, the probability that the treatment 
effect is > 10 units. Second, the probability that the 
treatment effect is > 5 units. Based on the results of these 
calculations, we need to decide what to do next and write the
R code accordingly.

- If the first probability is > 0.95, then print "Early 
  Success" and do nothing more. We do not have to run the 
  trial for the remaining patients.
- If the second probability is < 0.05, then print "Early 
  Futility" and do nothing more. We do not have to run the 
  trial for the remaining patients.
- Else, add 100 new patients to each group (200 overall) as 
  planned and analyze the full dataset with the 400 overall 
  patients using the JAGS model. 

At the final analysis for the trial, Provide R code to 
compute the posterior probability that the treatment 
difference is greater than 5 units.


\end{verbatim}

\textbf{BACTA-GPT Response 5}
\begin{verbatim}
# Generate the dataset for the interim analysis
# Generate age
age <- rnorm(200, mean = 8, sd = 4)
age <- pmax(age, 3)  # Truncate age at 3

# Generate treatment and outcome
X <- rep(c(0, 1), each = 100)
alpha <- 1.1  # True value of alpha
A <- age - mean(age)  # Center age
mu <- 10 + 6*X + alpha^A
Y <- rnorm(200, mean = mu, sd = 20)

# Create the dataset
trial_data <- data.frame(
  Y = Y,
  X = X,
  A = age
)

# Save the dataset
write.csv(trial_data, "clinical_trial_data_interim.csv", row.names = FALSE)

# Display summary statistics
summary(trial_data)

#Step 2: Restate the JAGS model
model_string <- "
model {
# Likelihood
for (i in 1:n) {
Y[i] ~ dnorm(mu[i], tau)
mu[i] = beta0 + beta1*X[i] + alpha^A[i]
}

# Priors
beta0 ~ dnorm(0, 1.0E-6)
beta1 ~ dnorm(0, 1.0E-6)
alpha ~ dunif(0, 1.5)
tau ~ dgamma(0.001, 0.001) # precision
sigma2 <- 1 / tau
}
"

#Step 3: Set up the environment and load libraries
library(rjags)
library(coda)

#Step 4: Run the model for the interim analysis
# Load the data
trial_data <- read.csv("clinical_trial_data_interim.csv")

# Prepare the data for JAGS
jags_data <- list(
  Y = trial_data$Y,
  X = trial_data$X,
  A = trial_data$A,
  n = nrow(trial_data)
)

# Initialize the model
jags_model <- jags.model(textConnection(model_string), 
                         data = jags_data, n.chains = 3)

# Burn-in
update(jags_model, 5000)

# Draw samples
mcmc_samples <- coda.samples(jags_model, 
                             variable.names = c("beta0", 
                             "beta1", "alpha", "sigma2"),
                             n.iter = 10000)
                             
# Summarize the results
print(summary(mcmc_samples))

# Extract the posterior samples
mcmc_list <- as.mcmc.list(mcmc_samples)

# Calculate the treatment difference for each iteration
treatment_diff <- as.matrix(mcmc_list[,'beta1'])

# Calculate the posterior probability that the treatment 
# difference is greater than 10
prob_treatment_diff_gt_10 <- mean(treatment_diff > 10)
prob_treatment_diff_gt_5 <- mean(treatment_diff > 5)

if (prob_treatment_diff_gt_10 > 0.95) {
  cat("Early Success\n")
} else if (prob_treatment_diff_gt_5 < 0.05) {
  cat("Early Futility\n")
} else {
  # Generate the full dataset
  # Generate age
  age <- rnorm(400, mean = 8, sd = 4)
  age <- pmax(age, 3)  # Truncate age at 3
  
  # Generate treatment and outcome
  X <- rep(c(0, 1), each = 200)
  alpha <- 1.1  # True value of alpha
  A <- age - mean(age)  # Center age
  mu <- 10 + 6*X + alpha^A
  Y <- rnorm(400, mean = mu, sd = 20)
  
  # Create the dataset
  trial_data <- data.frame(
    Y = Y,
    X = X,
    A = age
  )
  
  # Save the dataset
  write.csv(trial_data, "clinical_trial_data_full.csv", row.names = FALSE)
  
  # Display summary statistics
  summary(trial_data)
  
  # Load the data
  trial_data <- read.csv("clinical_trial_data_full.csv")
  
  # Prepare the data for JAGS
  jags_data <- list(
    Y = trial_data$Y,
    X = trial_data$X,
    A = trial_data$A,
    n = nrow(trial_data)
  )
  
  # Initialize the model
  jags_model <- jags.model(textConnection(model_string), 
                           data = jags_data, n.chains = 3)
  
  # Burn-in
  update(jags_model, 5000)
  
  # Draw samples
  mcmc_samples <- coda.samples(jags_model, 
                               variable.names = c("beta0",
                               "beta1", "alpha", "sigma2"), 
                               n.iter = 10000)
                               
  # Summarize the results
  print(summary(mcmc_samples))
  
  # Extract the posterior samples
  mcmc_list <- as.mcmc.list(mcmc_samples)
  
  # Calculate the treatment difference for each iteration
  treatment_diff <- as.matrix(mcmc_list[,'beta1'])
  
  # Calculate the posterior probability that the treatment 
  # difference is greater than 5
  prob_treatment_diff_gt_5 <- mean(treatment_diff > 5)
  
  cat("Posterior probability that treatment difference is", 
      "greater than 5 units:", prob_treatment_diff_gt_5, "\n")
}
\end{verbatim}

\end{document}